\documentclass{iau} 
\usepackage{graphicx}

\title[New models for the evolution of CSPN] 
{New models for the evolution of central stars of planetary nebulae: Faster and Brighter}

\author[Marcelo M. Miller Bertolami]   
{Marcelo M. Miller Bertolami$^1$}

\affiliation{$^1$Instituto de Astrofísica de La Plata, CONICET-UNLP, Paseo del Bosque s/n,\\ (B1900FWA) La Plata, Argentina. email: {\tt mmiller@fcaglp.unlp.edu.ar}}

\pubyear{2017}
\volume{323}  
\jname{PNe: Multi-wavelength probes of stellar and galactic evolution}
\editors{Xiaowei Liu, Letizia Stanghellini, \& Amanda Karakas., eds.}
\begin{document}

\maketitle

\begin{abstract}

 The post-asymptotic giant branch (AGB) phase is arguably one of the
 least understood phases of the evolution of low- and intermediate-
 mass stars. The recent post-AGB evolutionary sequences computed by
 \cite{2016A&A...588A..25M} are at least three to ten times faster
 than those previously published by \cite{1994ApJS...92..125V} and
 \cite{1995A&A...299..755B} which have been used in a large number of
 studies. This is true for the whole mass and metallicity range. The
 new models are also $\sim $0.1\textendash0.3 dex brighter than the
 previous models with similar remnant masses. In this short article we
 comment on the main reasons behind these differences, and discuss
 possible implications for other studies of post-AGB stars or planetary
 nebulae.

\end{abstract}

\firstsection 
\section{Introduction}

 In the most simple picture, Planetary Nebulae (PNe) are formed by
 low- and intermediate-mass stars ($M_{\rm ZAMS} \sim$ 0.8 \textendash
 6 $M_{\odot}$) after the strong stellar winds at the end of the
 Asymptotic Giant Branch (AGB). At that point, stars contract and heat
 up, crossing the HR-diagram at constant luminosity becoming
 sufficiently hot and bright to ionize the previously ejected material
 (\cite[Shklovsky 1957]{1957IAUS....3...83S}, \cite[Abell \& Goldreich
   1966]{1966PASP...78..232A}, \cite[Paczy{\'n}ski
   1970]{1970AcA....20...47P}).  The formation and detectability of
 PNe depends strongly on the interplay between two different
 timescales.  The evolutionary timescale of the central star of the PN
 (CSPN), which provides the ionizing photons, and the dynamical
 timescale of the circumstellar material \cite[(Sch{\"o}nberner et
   al. 2007)]{2007A&A...473..467S}. In this manuscript we focus on the
 evolutionary timescales of the CSPNe. In particular, we will discuss
 the new results presented by \cite{2016A&A...588A..25M} from full
 stellar evolution computations of the post-AGB and CSPN phases. These
 new post-AGB models are based on state-of-the-art stellar evolution
 computations, they include an updated treatment of the AGB microphysics
 (radiative opacities and nuclear reaction rates) as well as of the
 macrophysics (convective boundary mixing and mass loss rates). The
 models have been calibrated and tested to reproduce several
 observables during the post-AGB and previous evolutionary
 phases; e.g. solar radius and depth of the convective zone, width
 of the upper main sequence, mass range of carbon-rich AGB stars,
 oxygen abundances of PG1159 stars, semiempirical initial-final mass
 relationship, C/O ratios of AGB stars and CSPNe ---see
 \cite{2016A&A...588A..25M} for details.

\begin{figure}[t!]
\begin{center}
\includegraphics[width=\textwidth]{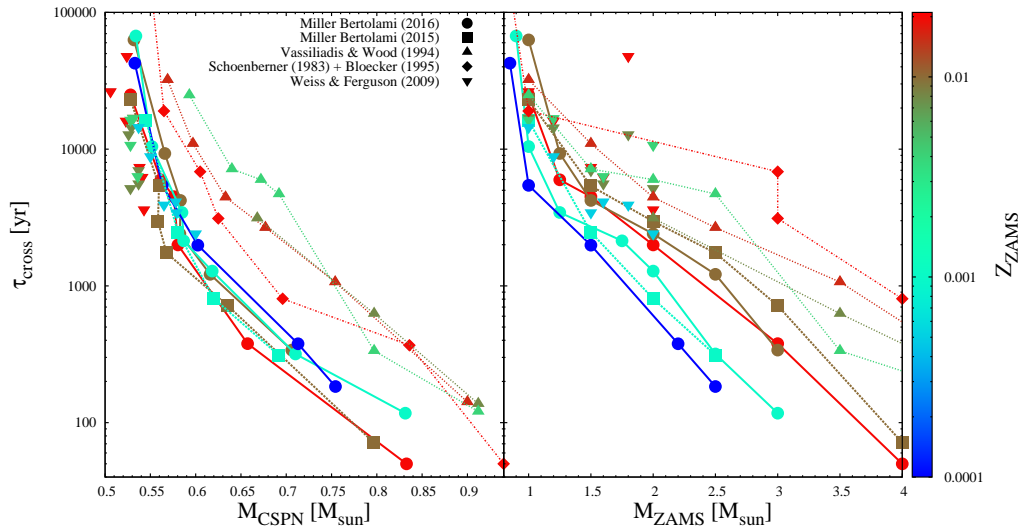}
\caption{Post-AGB crossing timescales ($\tau_{\rm cr}$) of the
  sequences presented by different authors. Following
  \cite{1994ApJS...92..125V} we have defined, and plotted, $\tau_{\rm
    cr}$ for all sequences as the time elapsed from $T_{\rm
    eff}=10000\,K$ to the point of maximum $T_{\rm eff}$ in the HR
  diagram during the post-AGB evolution. {\it Left Panel:} $\tau_{\rm
    cr}$ as a function of the final remnant mass $M_{\rm CSPN}$. {\it
    Left Panel:} $\tau_{\rm cr}$ as a function of the initial
  progenitor mass $M_{\rm ZAMS}$.  Color coding indicates the initial
  metallicities ($Z_{\rm ZAMS}$) of the sequences, as described in the
  color bar.}
\label{fig1}
\end{center}
\end{figure}

\begin{figure}[t!]
\begin{center}
\includegraphics[width=\textwidth]{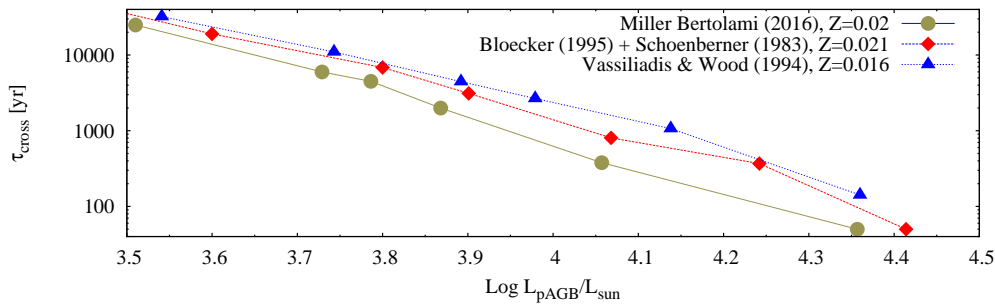}
\caption{$\tau_{\rm cr}$ as a function of the post-AGB luminosity for  sequences of similar metallicities.}
\label{fig2}
\end{center}
\end{figure}

\begin{figure}[t!]
\begin{center}
 \includegraphics[width=\textwidth]{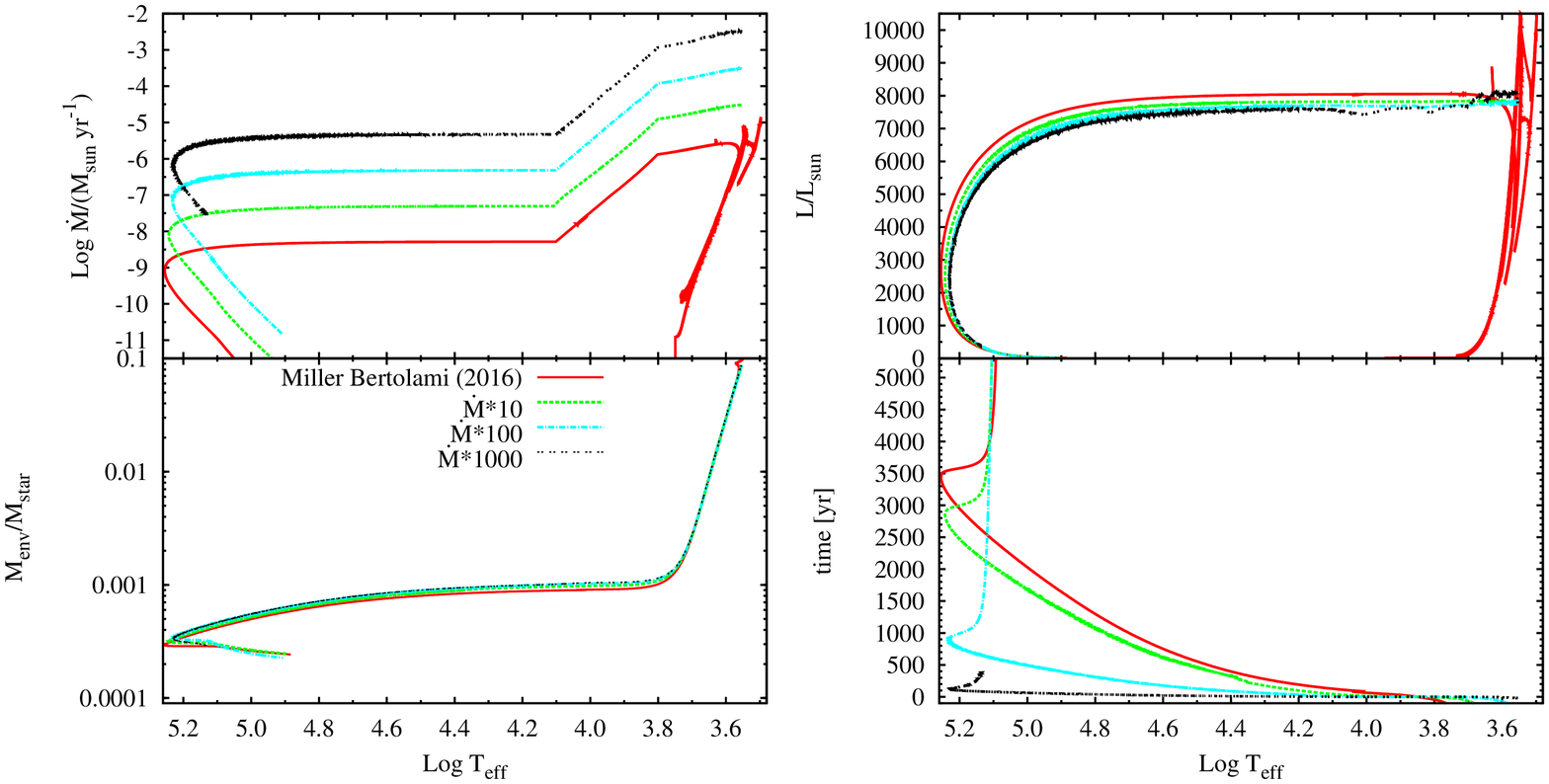} 
 \caption{A numerical experiment. Comparison of the late AGB and
   post-AGB evolution of sequences computed with winds differing by 3
   orders of magnitude. The reference sequence is the $M_{\rm
     ZAMS}=1.25M_\odot$, $Z=0.001$ sequence from
   \cite{2016A&A...588A..25M}. }
   \label{fig3}
\end{center}
\end{figure}

\section{New post-AGB/CSPN evolutionary timescales}
As shown in Fig.\ref{fig1}, the post-AGB/CSPN crossing timescales
($\tau_{\rm cr}$) computed by \cite{2016A&A...588A..25M} are 3 to 10
times shorter those of the sequences of similar remnant mass ($M_{\rm
  CSPN}$) computed by \cite{1994ApJS...92..125V} and
\cite{1995A&A...299..755B}. The new post-AGB models also evolve faster
when compared at equal initial masses ($M_{\rm ZAMS}$; see
Fig. \ref{fig1}) or post-AGB luminosities ($L_{\rm pAGB}$; see
Fig. \ref{fig2}).  The faster evolution of the new models can be
traced back to the adoption of updated microphysics and the inclusion
of convective boundary mixing during the thermal pulses on the AGB
---see \cite{2016A&A...588A..25M} for a detailed discussion. The
latter, in particular, leads to the occurrence of third dredge up for
lower stellar masses and to carbon (C) pollution of the envelope. In
fact, post-AGB timescales become even shorter in the case of the stellar
models computed by \cite{2009A&A...508.1343W} and
\cite{2015ASPC..493...83M}, which assume a more intense convective
boundary mixing during the thermal pulses on the AGB, and consequently
have stronger third dredge-up episodes.  C-pollution of the envelope
increases the luminosity of the burning shell and decreases the
critical envelope mass ($M^{\rm crit}_{\rm env}$) at which the models
depart from the AGB. Then, as $\tau_{\rm cr}$ is mostly determined by the
pace at which the hydrogen(H)-burning shell consumes the remaining
H-rich envelope, an increase in the C-pollution of the envelope leads to a
faster post-AGB evolution. It is worth emphasizing that the value of
$M^{\rm crit}_{\rm env}$ cannot be arbitrarily set for a given
sequence. $M^{\rm crit}_{\rm env}$ is mostly set by the H-burning
shell luminosity as well as by the composition of the H-rich
envelope. As noted by \cite{1971AcA....21..417P}, H-burning post-AGB
models have a very tight relationship between the effective
temperature ($T_{\rm eff}$) and the envelope mass ($M_{\rm env}$). The
departure from the AGB occurs as soon as $M_{\rm env}$ is reduced
close to these corresponding post-AGB values. In particular, this
means that the wind intensity at the end of the AGB plays no role in
the determination of $M^{\rm crit}_{\rm env}$. This is fortunate, as
the intensity of $\dot{M}$ at the end of the AGB, or the early
post-AGB, is currently not known. In order to convince the reader of
this counter-intuitive fact, we have recomputed the $M_{\rm
  ZAMS}=1.25M_\odot$, $Z_{\rm ZAMS}=0.001$ sequence presented by
\cite{2016A&A...588A..25M} artificially increasing the mass loss
($\dot{M}$) after the last thermal pulse up to 3 orders of magnitude
(Fig. \ref{fig3}, upper left panel). As seen in Fig. \ref{fig3}
(bottom left panel) the value of $M^{\rm crit}_{\rm env}$ is
independent of $\dot{M}$ at the end of the AGB. Only the post-AGB
timescales are affected (Fig. \ref{fig3}, bottom right panel) because
at high values of $\dot{M}$ the H-rich envelope is mostly reduced by
winds, speeding up the evolution. It should be noted, however, that
$\dot{M}_{\rm AGB}$ can affect the post-AGB envelope mass in an
indirect way.  As $M^{\rm crit}_{\rm env}$ is strongly dependent on
the luminosity of the sequence, and luminosity does not stay constant
during the AGB phase, different $\dot{M}$-values during the whole AGB
evolution can change the value of $M^{\rm crit}_{\rm env}$ for a given
sequence.

\section{Implications and final comments}

Given the significantly shorter timescales, higher luminosities and
different initial-final mass relations of the models presented by
\cite{2016A&A...588A..25M}, as compared with the usually adopted
tracks from \cite{1994ApJS...92..125V} and \cite{1995A&A...299..755B},
one may wonder about the  impact of these results for studies
that rely on stellar evolution models as inputs. For example,
\cite{2016arXiv160908680G} studied 32 PNe from the bulge and obtained
CSPNe-black-body temperatures and PNe expansion ages. They find that
the CSPNe and progenitor masses derived from the post-AGB models of
\cite{2016A&A...588A..25M} are in agreement with our current
understanding of the stellar formation history of the Galactic Bulge
and the white dwarf mass distribution. This was not the case when the
tracks of \cite{1994ApJS...92..125V} and \cite{1995A&A...299..755B}
were used \cite[(Gesicki et al. 2014)]{2014A&A...566A..48G}.

The fact that, for $M_{\rm ZAMS}\gtrsim 1.25\, M_\odot$, the post-AGB
timescales of the new models are shorter than those of
\cite{1994ApJS...92..125V} and \cite{1995A&A...299..755B} for the same
initial masses (Fig. \ref{fig1}, right panel) will have an impact in
the understanding and modeling of the Planetary Nebula Luminosity
Function (PNLF). \cite[Mendez (2016)]{2016arXiv161008625M} concludes
that with the new stellar models, CSPNe with masses as low as $0.58\,
M_\odot$ are able to create relatively bright PNe ---see also Zijlstra
et al. in these proceedings. Consequently, the progenitors of PNe
close to the cut-off of the PNLF will be older than previously
expected, something that will help to explain the lack of sensitivity
of the PNLF cut-off to the age of the harboring population. Due to
their shorter post-AGB timescales and larger brightnesses, the new
models are also expected to have an impact on the study of the diffuse
X-ray emission from the inner regions of planetary nebulae by means of
radiation-hydrodynamics numerical simulations
---e.g. \cite{2008A&A...489..173S} and \cite{2014MNRAS.443.3486T}.  In
addition to these results, we can also speculate that the shorter
post-AGB timescales of the new models will help to understand the
formation PNe around low mass post-AGB stars ---e.g. as determined by
asteroseismology, see \cite{2008A&A...478..175A},
\cite{2016A&A...589A..40C} and references therein. Also, shorter
post-AGB timescales might help to understand the lack of CSPNe and
post-AGB stars in M32 \cite[(Brown et al. 2008)]{2008ApJ...682..319B}.

A word of caution, while the new models are to be preferred over the
older grids, they are not devoid of uncertainties. In particular, the
stellar models of \cite{2016A&A...588A..25M} fail to quantitatively
reproduce the lifetimes of M- and C-type AGB stars, pointing to a need
of a better calibration of convective boundary mixing and AGB-winds.

{\it Acknowledgments:} M3B thanks the IAU and the organizers for a
travel grant and the waiving of the registration fee which allowed him
to attend this wonderful Symposium. Part of this work was supported by ANPCyT
through grant PICT-2014-2708 from FonCyT and grant PIP
112-200801-00940 from CONICET and by a Return Fellowship from the
Alexander von Humboldt Foundation.


\end{document}